\documentclass[12pt]{article}

\usepackage{graphicx}
\usepackage{fullpage,amsmath,amssymb,latexsym}
\usepackage{multirow}
\usepackage{rotating} 
\usepackage{amsmath}
\usepackage{url}
\usepackage{graphicx}
\usepackage{caption}
\usepackage{subcaption}

\begin{document}

\title{Methane Planets and their Mass-Radius Relation}
\author{Ravit Helled, Morris Podolak \& Eran Vos\\
Department of Geosciences, 
Tel-Aviv University, Tel-Aviv, Israel.}
\date{}
\maketitle

\begin{abstract}
Knowledge of both the mass and radius of an exoplanet allows us to estimate its mean density, and therefore its composition. Exoplanets seem to fill a very large parameter space in terms of mass and composition, and unlike the solar-system's planets, exoplanets also have intermediate masses ($\sim$5 - 50 M$_{\oplus}$) with various densities. 
In this letter, we investigate the behavior of the Mass-Radius relation for methane (CH$_4$) planets and show that when methane planets are massive enough (M$_p \gtrsim$~15~M$_\oplus$), the methane can dissociate and lead to a differentiated planet with a carbon core, a  methane envelope, and a hydrogen atmosphere. The contribution of a rocky core to the behavior of CH$_4$ planet is considered as well. We also develop interior models for several detected intermediate-mass planets that could, in principle, be methane/methane-rich planets. 
The example of methane planets emphasizes the complexity of the Mass-Radius relation and the challenge involved in uniquely inferring the planetary composition.  
\end{abstract}

\date{}
\maketitle
\section{Introduction}
Exoplanet studies are now reaching the level at which planet characterization is possible. 
There are hundreds of planets with measured mass and radius, not including the thousands of candidates from the Kepler mission. 
Characterization of exoplanets is often performed via the Mass-Radius relationship (hereafter, M-R relation) which provides information on the planet's mean density, and therefore its composition. The curves of planetary radius versus mass for various compositions have been computed by several groups (e.g., Valencia et al., 2006; Sotin et al., 2007; Seager et al. 2007; Fortney et al., 2007; Rogers et al., 2011; Lopez \& Fortney 2014; Lissauer et al., 2011; Fressin et al., 2012; Wagner et al., 2011).  
The increasing number of observed exoplanets with masses of 5 - 20 M$_\oplus$ and the large variation in their mean densities encourages the investigation of compositions that include volatile materials and various mixtures.   
\par

Typically, the M-R curves are smooth and monotonic.  Discontinuities in the M-R curve could, however, appear when phase changes are considered.  An example of this situation can occur in hydrogen-dominated planets, where at pressure of $\sim$100 GPa molecular hydrogen transitions to metallic hydrogen. This transition leads to a kink in the M-R curve for planets with masses of 100 M$_{\oplus}$ (e.g., Swift et al., 2012). 
\par

Discontinuities in the M-R relation can also occur for other materials. One example is methane (CH$_4$). Here, we show that CH$_4$ planets can dissociate and lead to a formation of a carbon (diamond) core, a CH$_4$ envelope, and an H$_2$ atmosphere. Under such a configuration, the planetary radius can be significantly larger than that expected without dissociation, which can affect the composition inferred from the M-R relation. CH$_4$ dissociation leading to a formation of a carbon core has already been presented in the context of the solar-system giant planets (Ross, 1981; Benedetti et al., 1999). 
Similar behavior can occur for other molecules, but in this paper we investigate only CH$_4$. Methane is a good example because this molecule is expected to be abundant in planetary systems. In the solar-system, for example, all of the outer planets are found to be enriched with carbon (by measuring the carrier CH$_4$) with Jupiter and Saturn being enriched with carbon by factors of about 4.3 and 9.7 compared to the proto-solar value, respectively, and Uranus and Neptune by factors of about 86 and 68, respectively (e.g., Guillot \& Gautier, 2014), although these numbers correspond only to the atmospheric composition and it is unclear whether they represent the bulk compositions of the planets. Clearly, other molecules such as H$_2$O, NH$_3$, and PH$_3$ could exist as well, and in fact, a planet that is enriched with carbon is likely to have much oxygen as well. However, modeling CH$_4$ planets is already non-trivial due to the possibility of dissociation and differentiation. 

Finally, while the behavior of refractory materials such as silicon and iron is relatively insensitive to temperature, this is not the case for volatile materials for which could increase the planetary radius  considerably with increasing temperature.  Below, we explore how the choice of the dissociation pressure and internal temperature affect the M-R relation for methane planets, and investigate how the inclusion of a silicate (or iron) core changes the results. Finally, we identify several detected exoplanets that could be methane-rich planets. 
\par

\section{Modeling Methane Planets}

\subsection{Computing the internal structure}
The internal structure is given by the basic equations of planetary structure: the equation of hydrostatic equilibrium and the equation of mass conservation.  For simplicity, an energy transport equation is not included and the planets are assumed to be isothermal. 

The structure equations are integrated using a fourth-order Runge-Kutta procedure. We integrate from a central pressure $P_c$ to a low surface pressure $P_s$ (typically set to 1 bar) and compute the planetary mass and radius. For pure CH$_4$ planets, if $P_c$ is greater than the methane dissociation pressure, hereafter, $P_{\rm diss}$, we use the equation of state (EOS) for pure carbon.  The integration is continued until the pressure falls below $P_{\rm diss}$.  The mass of the body at this point is the mass of the carbon core, $M_C$.  The dissociation of CH$_4$ produces an additional $1/3M_C$ of hydrogen which we assume has diffused to the outer envelope of the planet to form a hydrogen atmosphere.  Any remaining mass is in the form of an intermediate shell of CH$_4$.

In principle, each planetary mass should correspond to a unique $P_c$ value, and we can find a single-valued function $M(P_c)$ for the planetary mass versus central pressure. However this is not always the case. In the immediate vicinity of $P_c=P_{\rm diss}$, central pressures just below and just above $P_{\rm diss}$ give the same mass (e.g., Ramsey, 1948). In this case, the planet with a  carbon core is the stable configuration.  

For planets with a silicate/iron core (see section 2.4) we follow the same procedure, except that the innermost region is assumed to be SiO$_2$/Fe.  We integrate out from the center until the mass of the body equals the mass of the prescribed silicate/iron core. After that, if the pressure is greater than $P_{\rm diss}$ we use the equation of state for carbon, otherwise we use the EOS for CH$_4$.  Also here we guarantee that the mass of the hydrogen atmosphere is one-third the mass of the carbon shell.

\subsection{The equation of state}
We use the SESAME EOS for carbon, methane, and hydrogen (e.g., Kerley, 1972). When cores are included, we use the SESAME EOS for SiO$_2$ and iron (Fe). 
In order to account for the effect of methane dissociation, we must assume the pressure at which dissociation occurs. Several studies suggest that methane dissociates at a pressure of $\sim$300 GPa (Chau et al., 2011; Gao et al., 2010; Ancilotto et al., 1997), thus, the exact dissociation pressure of CH$_4$ is still unknown and needs to be confirmed through experiments and theoretical calculations.  In order to investigate the 
sensitivity of the derived internal structure (and M-R relation) to the dissociation pressure, we also consider a lower dissociation pressure of 170 GPa. 
The lower the dissociation pressure is, the lower the planetary mass for which a jump occurs in the M-R relation. In fact, before CH$_4$ breaks down to carbon and hydrogen some intermediate products of hydrocarbons can be formed: methane will first form ethane (C$_2$H$_6$) at 95 GPa, butane (C$_4$H$_{10}$) at 158 GPa, and finally carbon (diamond) and hydrogen (Gao et al., 2010). 

If hydrogen diffuses through the shell of undissociated CH$_4$ to the planetary atmosphere, then pure CH$_4$ planets above a certain mass would have a pure carbon core, a methane inner shell, and a (molecular) hydrogen atmosphere. In this work, we assume that once dissociation occurs differentiation takes place. 
Modeling the actual of differentiation is beyond the scope of this paper. Some pioneering work in that direction was recently presented by Levi et al.~(2014).  
 
\subsection{Pure methane planets}
For a pure methane planet, there is a sharp discontinuity in the M-R relation above the mass where dissociation occurs. This is presented in the two top panels of  Figure 1.  Since unlike for refractory materials, the temperature is not negligible, we consider four different temperatures: 50 K, 500 K, 1,000 K, and 1,500 K. 
The isotherm of 50K should be taken as extreme-low-temperature while other isotherms are comparable to the atmospheric temperatures of close-in exoplanets. Of course, planets are not isothermal but are expected to have a temperature profile, with the temperature increasing toward the center. Thus, since the temperature mostly affects the planetary radius when dissociation occurs due to the formation of an hydrogen envelope, assuming isothermal structure is reasonable. The effect of temperatures is much less profound in the deep interior that is composed of refractory materials. In addition, the atmospheres of close-in planets are expected to be nearly isothermal down to a depth of $\sim$ 100 bar (e.g., Madhusudhan \& Seager,  2010), the bottom of the hydrogen envelope is convective, in this case an adiabat is more appropriate but since the region between 100 bar and the hydrogen-methane boundary is relatively small an adiabat profile for the temperature only increases the temperature by a factor of $\sim$ three. In fact, since in our case the atmosphere is composed of pure hydrogen the opacity is expected to be lower than the typically modeled atmospheres since they include H$_2$O, CO, CH$_4$ and CO$_2$ that increase the opacity, and therefore the radiative-convective boundary of pure-hydrogen atmospheres, as in our case, will be at even higher pressures. We can then conclude that an isothermal structure is a reasonable assumption but clearly the sensitivity of the assumed temperature profile on the structure should be investigated in the future. The effect of temperatures on the M-R relation is strongest in the outer region (hydrogen) and is much less profound in the deep interior that is composed of refractory materials.  
\par

The derived M-R relations are shown for $P_{\rm diss}$ of 170 GPa (left) and 300 GPa (right). The solid, dashed, dotted, and dashed-dotted curves correspond to T = 50 K, 500 K, 1,000 K and 1,500 K, respectively. As long as the planet is undifferentiated, the radius is insensitive to temperature. When the planetary mass is high enough to produce a hydrogen (H$_2$) atmosphere, the radius increases abruptly and becomes much more sensitive to temperature. 
The amount of methane in the planet varies with its mass.  As long as the planetary mass is low enough so that dissociation does not occur, the mass fraction of methane is unity and the radius is insensitive to the temperature, thus, once the critical mass (pressure) for dissociation is reached, the mass fraction of methane decreases rapidly. 
\par

Hot methane planets that dissociate and develop a molecular hydrogen envelope, could, in principle, lose some of their atmospheres. 
For a sun-like star, the four temperatures of 50 K, 500 K, 1,000 K and 1,500 K correspond to radial distances of about 30 AU, 0.33AU, 0.083 AU and 0.037 AU, respectively. 
The most vulnerable planets are those with the high temperatures (i.e., T = 1,000 K and T = 1,500 K). Planets with small masses have $P_c~<~P_{\rm diss}$ and therefore are not expected to have hydrogen atmospheres.  For the cases we investigate, this means masses less than $\sim8$~M$_{\oplus}$ (see Fig.\,1).  More massive planets can dissociate CH$_4$, but for these masses hydrogen escape is less efficient due to the relatively large gravity. 
We can estimate the mass-loss rate for the different cases by using the energy-limited mass loss estimates (e.g.,  LeCavelier des Etangs, 2007; Lopez et al., 2012, Owen \& Jackson 2012; Murry-Clay et al., 2009; Mordasini et al., 2015; Kurosaki et al., 2014).  
For planets with temperatures of 50 and 500 K evaporation is found to be negligible. For temperatures of 1,000 K and 1,500 K, planets with masses larger than $\sim$~7~M$_{\oplus}$ the mass that is lost after 10 Gyr is about 2 \% and 10 \% of the total planetary 
mass, respectively\footnote{The estimates were performed assuming energy-limited mass loss  around a G star, and accounting for the change in the X-ray and UV radiation flux at stellar age of 0.1 Gyr (Ribas et al., 2005) integrated up to an age of 10 Gyr. More information can be found in Kurosaki et al.~(2014) and references therein. These estimates however are very crude since the change in internal structure due to evaporation is not included.}. 
Smaller planets, however, can lose a significant amount of mass, especially if they are close to their host star. Therefore, low-mass planets with hydrogen atmospheres are unlikely to exist in old systems. Thus, the mass loss estimate assumes a pure-hydrogen atmosphere, and since small planets cannot dissociate (due to low internal pressures), they have methane atmospheres, for which the mass loss rate can be much lower. Further investigations of the mass loss rates of various atmospheric compositions is required to understand the survivability of such planets (e.g., Lammer et al., 2012).

\subsection{The effect of a silicate core}
Since planets are not expected to consists of pure methane, we next consider the existence of a silicate (SiO$_2$) core and investigate how it affects the derived M-R relation.  
The presence of a silicate core reduces the importance of this effect.  As the silicate mass fraction of the planet increases, the magnitude of the radius change decreases.  This is because the silicate core occupies part of the high pressure region where the methane dissociation should take place.  The larger the silicate core, the smaller the volume of methane that participates in the dissociation, and the smaller the difference in radius at the transition point. The effect of the silicate core is demonstrated in the bottom panels of Figure 1 where we compare the M-R relation for methane planets with no core (solid-black) and with silicate cores that are equal to 30\% (dotted-gray), 50\% (dashed-blue) and 80\% (dashed dotted-red) of the total planetary mass for T= 500 K for $P_{\rm diss}$ of 170 GPa (left) and 300 GPa (right). 
While we use SiO$_2$ to represent the core material, the core could be composed of other elements such as Fe or MgO. The results are not expected to change much qualitatively since the effect is similar as long as the core material is significantly denser than methane. A comparison of the inferred internal structure for an SiO$_2$ core vs.~an Fe core is discussed in the following section.  
As can be seen from the bottom panels of Figure 1, the discontinuity due to CH$_4$ dissociation is much less prominent when the planet consists of a massive core. 
When the silicate core consists of 30\% and 50\% of the total planetary mass, dissociation occurs at smaller masses. This is because the core results in a higher internal pressure which allows dissociation at lower masses than in the no-core case. Thus, at some point, when the core mass is significantly larger than the methane mass, methane will exist only at the outer region (envelope) where there pressure is low. In that case, dissociation occurs at higher masses compared to the no-core case. Indeed, for the case of 80\% core dissociation is found to occur at higher masses. Here too the kink in the M-R relation occurs for larger masses for the higher $P_{\rm diss}$.  

\section{Fits to observed exoplanets} 
Several detected exoplanets have relatively low masses but radii that are larger than expected from pure H$_2$O, and therefore must consist of an atmosphere. Here, we show that some of these planets are good candidates for methane-dominated planets. In Figure 2 we show the M-R relation for planets with masses up to 20 M$_{\oplus}$ for various materials including several detected exoplanets. The curves for pure iron (brown), 100\% MgSiO$_3$ (black), H$_2$O-MgSiO$_3$ (green), and 100\% water (cyan) are taken from  Zeng \& Sasselov (2013). Also shown are the M-R curves for methane/methane-rich planets. 
Several of the detected exoplanets have, in principle, masses and radii that are consistent with those predicted for methane/methane-rich planets. 
Clearly, there are other possible solutions and as argued in this Letter, there is no unique solution even for methane planets. Nevertheless, the possibility of methane-dominated planets is intriguing. 

Examples for potential-methane planets are GJ3470b and Kepler 89c\footnote{The estimate for the mass and radius used in this paper for Kepler-89c were based on Weiss et al.~(2013) with M = 15.57$_{-5.72}^{+15.57}$ M$_{\oplus}$ and R = 4.33 $\pm$ 0.09 R$_{\oplus}$. New estimates for the mass and radius of the planet were derived by Masuda et al.~(2013), who suggest a planetary mass of 9.4$_{-2.1}^{+2.4}$ M$_{\oplus}$ and radius of 3.8$ _{0.26}^{+0.29}$ R$_{\oplus}$. If Kepler-89c is indeed less massive and smaller as suggested by Masuda et al.~(2013), Kepler-89c will not be able to differentiate for a dissociation pressure of 300 GPa, and in addition, for the cases were differentiation does occur (170 GPa) the internal structure of the planet will change as well. For the numbers suggested by Masuda et al.~(2013) a smaller fraction of methane can dissociate, and therefore, Kepler-89c will have smaller carbon and molecular hydrogen shells than presented here. Thus, Kepler-89c still remains a potential differentiated-methane-rich planet for $P_{\rm diss}$ = 170 GPa.}. The inferred internal structure for these planets are shown in Figure 3 (left panel). The inferred interior structure is shown for four different temperatures: 50 K (blue), 500 K (gray) 1,000 K (black), and 1,500 K (red). For colder planets, the mass fraction of the silicate core is larger, but in all the cases the planet has a differentiated structure. For Kepler-89c, the region of the methane shell is found to be significantly larger for $P_{\rm diss}$ = 300 GPa, and the hydrogen atmosphere is more compact. In order to demonstrate the sensitivity of the internal inferred structure to the core material we compare the solutions for T = 1,000 K for SiO$_2$ and iron (Fe) cores (right panels). 

The derived values for the mass fractions and locations for the different elements are summarized in Table 1, also listed are other exoplanets for which there is a solution consistent with differentiated methane planets.  Examples for exoplanets that are consistent with an internal structure which consists of SiO$_2$ and CH$_4$ but with undissociated methane are listed in Table 2. 
 The uncertainty in the measured radii and masses can be very large. Here, we set the mass and radius of a given exoplanet to be their 0-sigma values. Clearly, the uncertainty in mass and radius introduce a large uncertainty on the actual internal structure of the planet. This emphasizes the need for accurate determinations of the masses and radii of exoplanets. 

\section{Discussion}
Observations of exoplanets have taught us that planets come in a variety of masses and compositions. 
While the M-R relation is used  for classifying exoplanets, the relation between mass and radius is not always straightforward.   
Here, we show that even a simple composition consisting of one species, such as methane, can lead to complications due to phase transition and differentiation. In addition, the temperature must be accounted for when the exoplanets consist of volatile materials. 
We show that some of the observed masses and radii of intermediate-mass exoplanets are, in principle, consistent with methane-rich interiors. 

Our model is rather simple. A necessary next step would be to derive M-R relations for planets with more realistic temperature profiles (e.g., an adiabatic temperature profile). In addition, future models should account for other possible mixtures. In particular, it is interesting to investigate how the presence of oxygen affects the derived internal profile, and to investigate whether or not  the existence of other elements prevents the formation of a hydrogen atmosphere.  
The presented research can be continued by accounting for more materials, but for that case chemical interactions should be modeled as well. 
Finally, a better theoretical understanding of the differentiation of methane planets and knowledge of the behavior of methane at high pressure and temperatures are required.  
Upcoming space missions such as TESS, CHEOPS and PLATO 2.0 and ground-based observations will provide new and exciting data, and it is now an ideal time to establish a deeper understanding of the M-R relation and its potential for characterizing exoplanets. 

\subsection*{Acknowledgments} 
We thank an anonymous referee for valuable comments. 
R.~H.~acknowledges support from the Israel Space Agency under grant 3-11485.

\clearpage

\begin{figure}[ht]
\centerline{\includegraphics[angle=0, width=16cm]{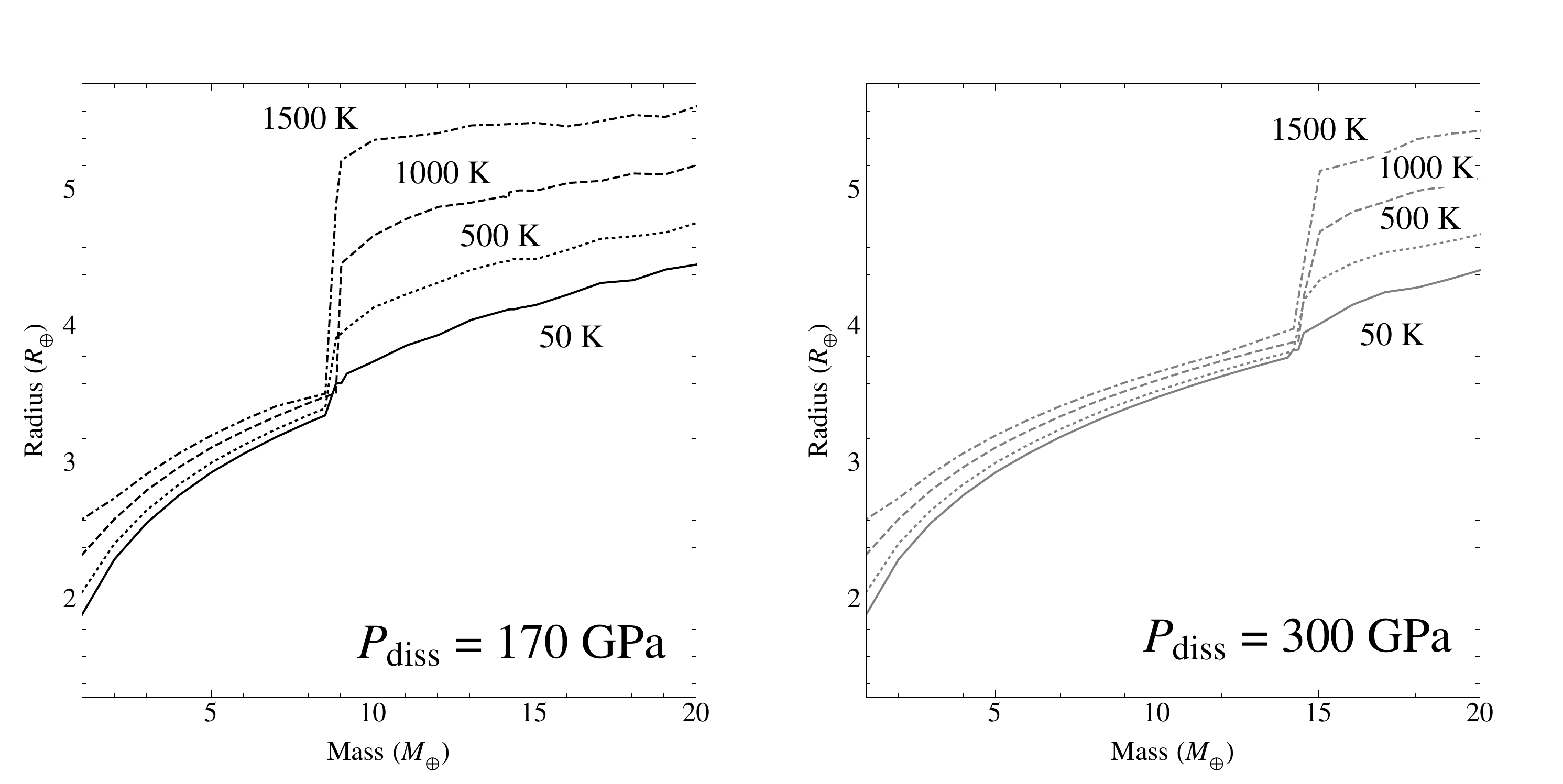}}
\centerline{\includegraphics[angle=0, width=16cm]{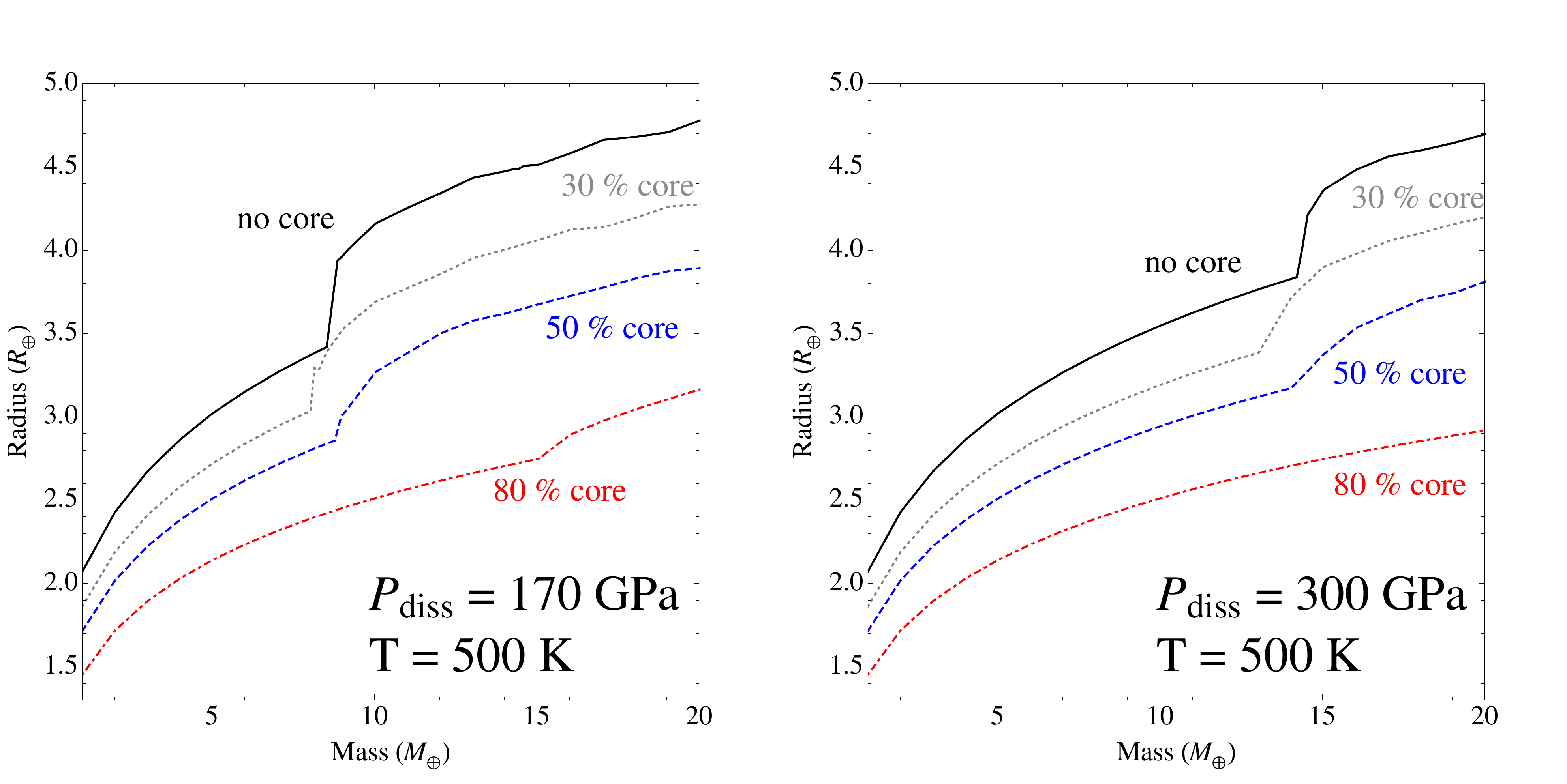}}
\caption{
  {\footnotesize {\bf Top}: M-R relation for pure CH$_4$ planets with isotherms of 50 K (solid), 500 K (dotted), and 1,000 K (dashed), and 1,500 K (dashed-dotted)  assuming a dissociation pressure of 170 GPa (left panel) and 300 GPa (right panel).
{\bf Bottom}: M-R relation for methane planets with different mass fractions of a silicate (SiO$_2$) core for a temperature of 500 K.  Also here the left and right panels correspond to $P_{\rm diss}$ of 170 GPa and 300 GPa, respectively. See text for details.}
}
\end{figure}

\begin{figure}[ht]
\centerline{\includegraphics[angle=0, width=15cm]{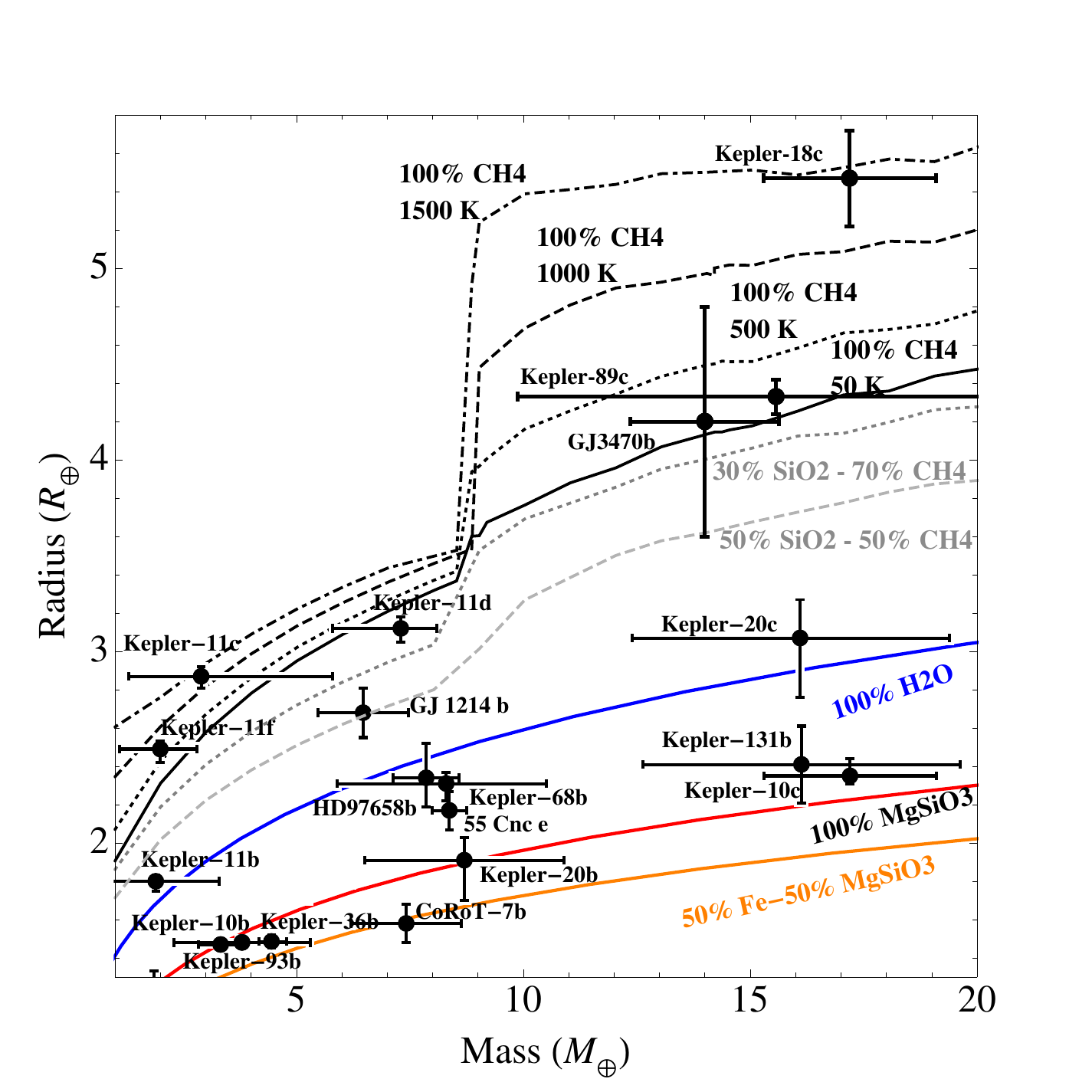}}
\caption{
  {\footnotesize M-R relation for various compositions including methane, and several exoplanets with detected masses and radii. The curves for H$_2$O, Fe-MgSiO$_3$, MgSiO$_3$ are taken from Zeng \& Sasselov (2013). The black curves correspond to pure CH$_4$ using $P_{\rm diss}$ = 170 GPa for the four temperatures we consider (similar to Fig.~1, top-left). The gray curves show the M-R curves for methane planets with silicate (SiO$_2$) cores consisting of 30\% (dotted) and 50\% (dashed) of the total planetary mass. Several planets are found to have masses and radii that are consistent with pure methane planets, such as Kepler 89c, GJ3470b, and three of the Kepler 11 planets: Kepler 11c, Kepler 11d, and Kepler 11f. The masses and radii of Kepler 11b and GJ 1214b fit a mixture of methane and rocks. }
  }
\end{figure}

\clearpage

\begin{figure}
\centering 
  \begin{subfigure}[b]{0.495\textwidth}
    \includegraphics[width=\textwidth]{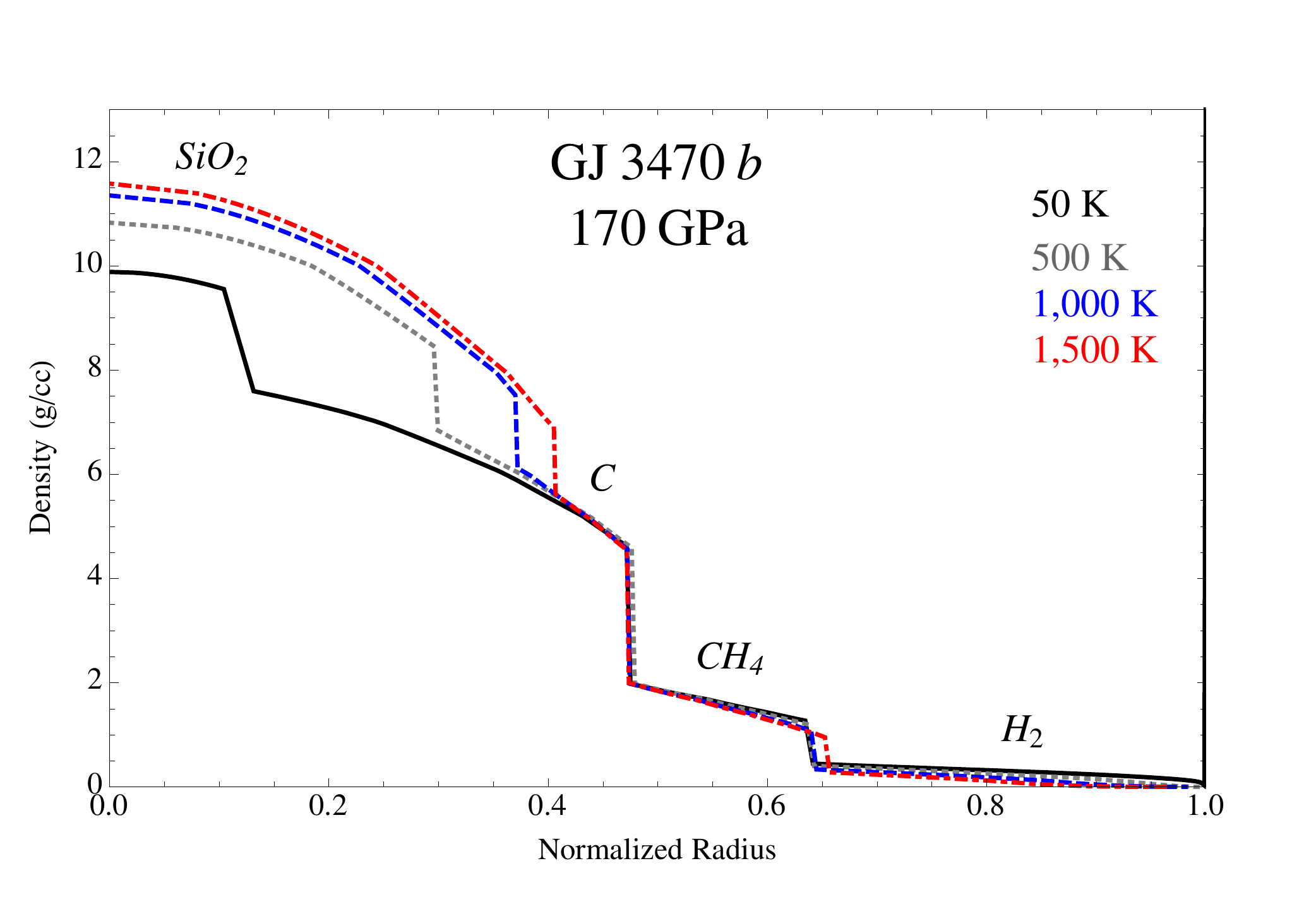}
    \label{fig:1}
  \end{subfigure}
  \vspace{-22pt}
  \begin{subfigure}[b]{0.495\textwidth}
    \includegraphics[width=\textwidth]{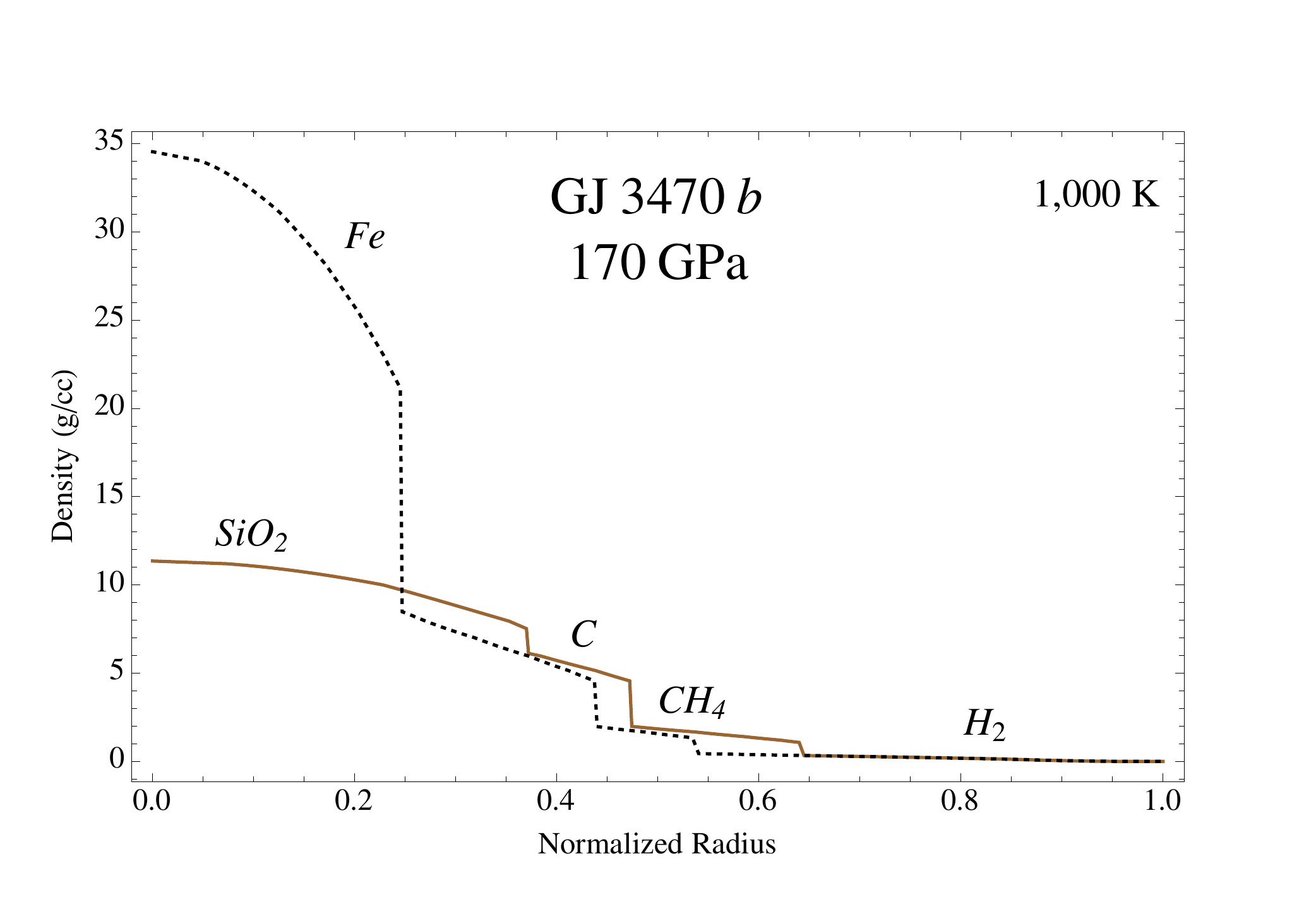}
    \label{fig:2}
  \end{subfigure}
    \begin{subfigure}[b]{0.495\textwidth}
    \includegraphics[width=\textwidth]{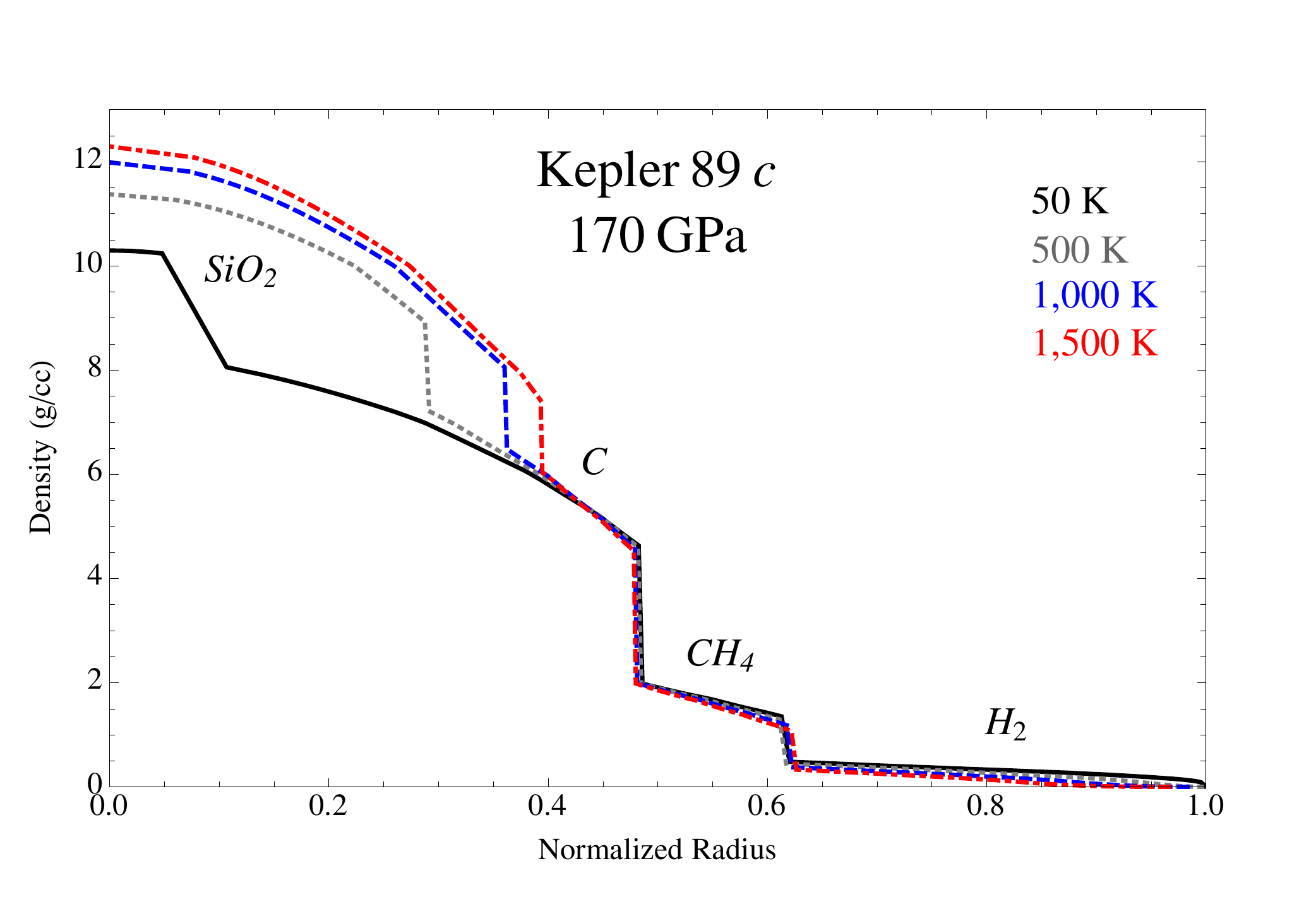}
    \label{fig:1}
  \end{subfigure}
  \vspace{-22pt}
  \begin{subfigure}[b]{0.495\textwidth}
    \includegraphics[width=\textwidth]{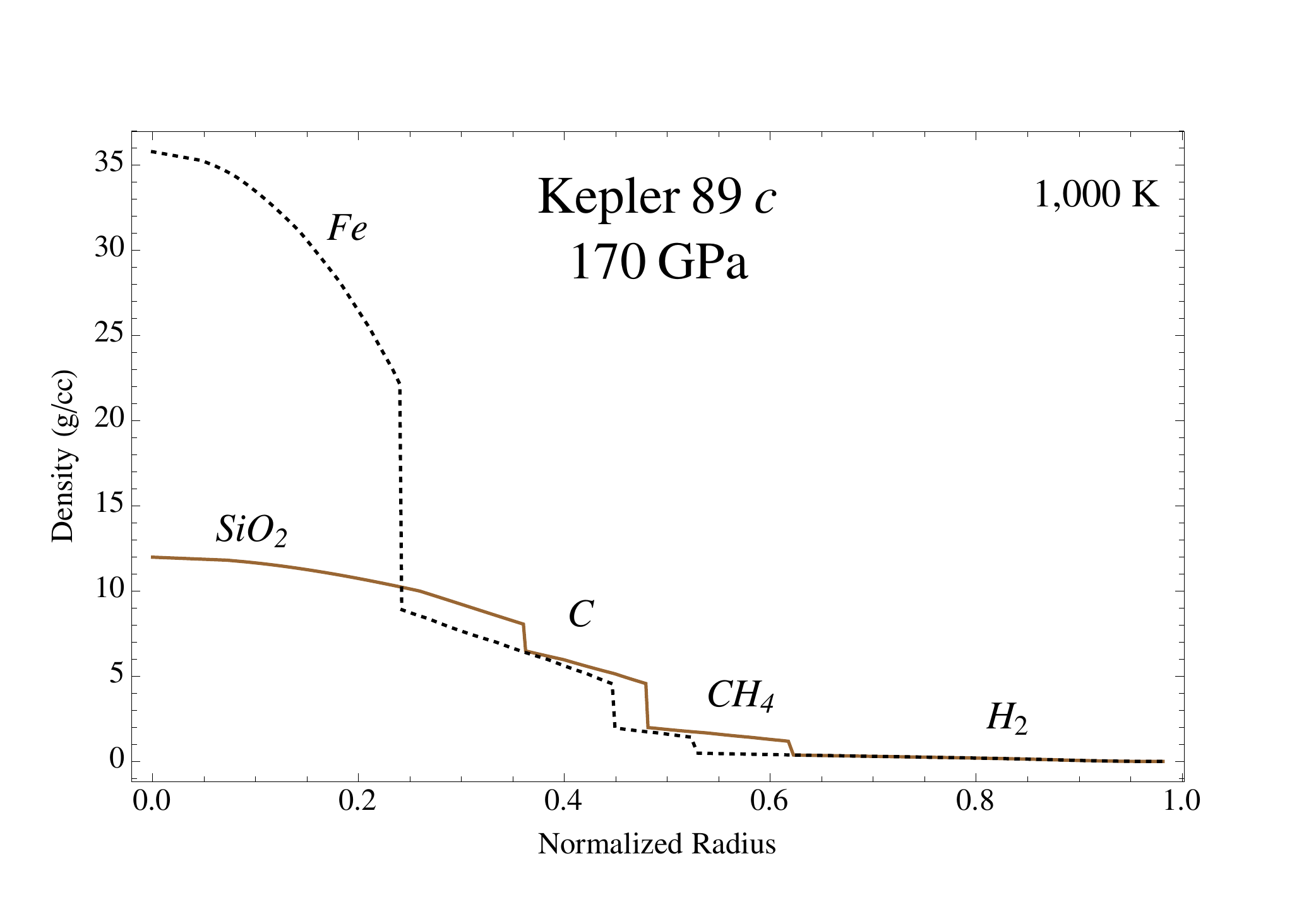}
    \label{fig:2}
  \end{subfigure}
    \begin{subfigure}[b]{0.495\textwidth}
    \includegraphics[width=\textwidth]{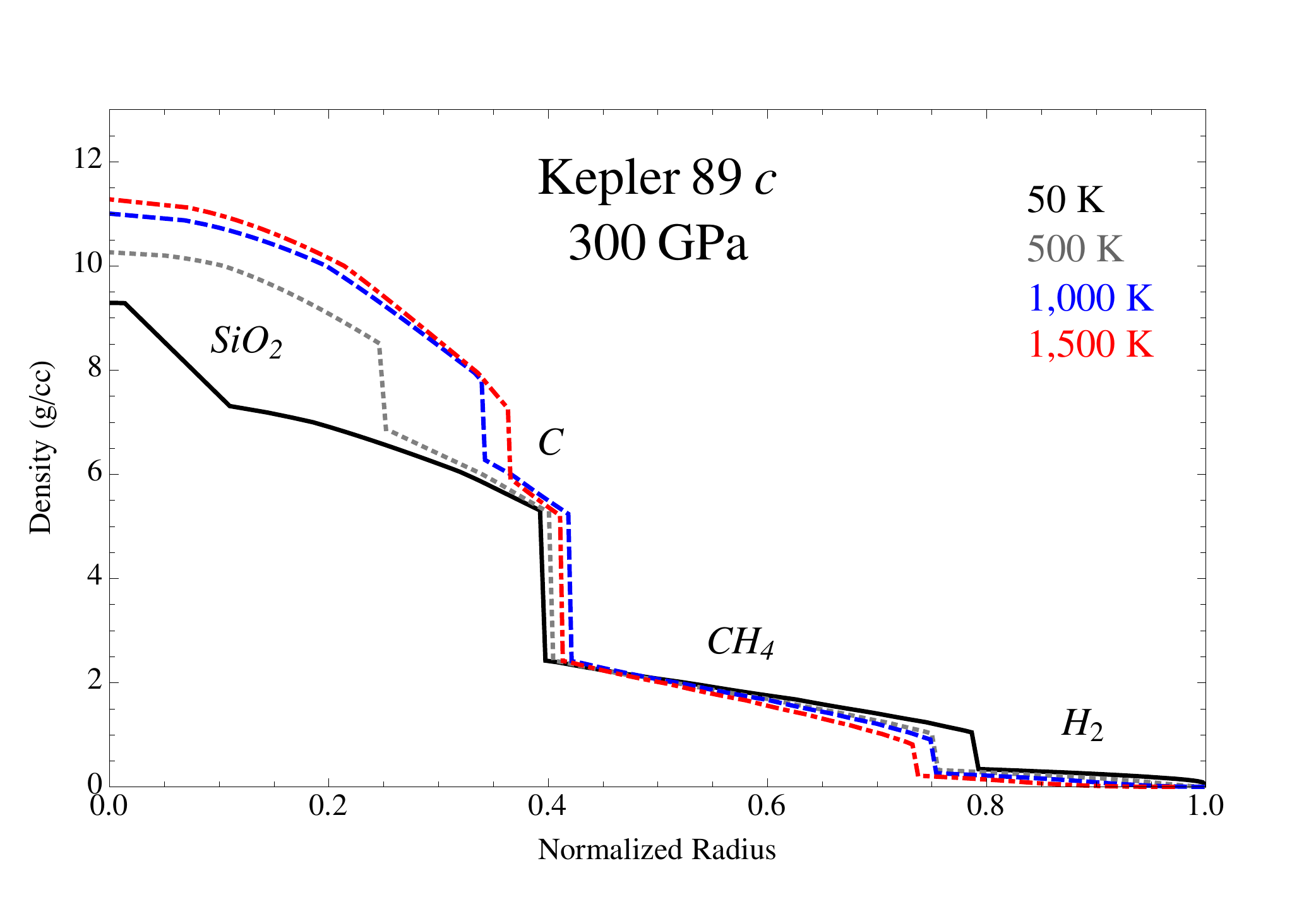}
    \label{fig:1}
  \end{subfigure}
  \vspace{-20pt}
  \begin{subfigure}[b]{0.495\textwidth}
    \includegraphics[width=\textwidth]{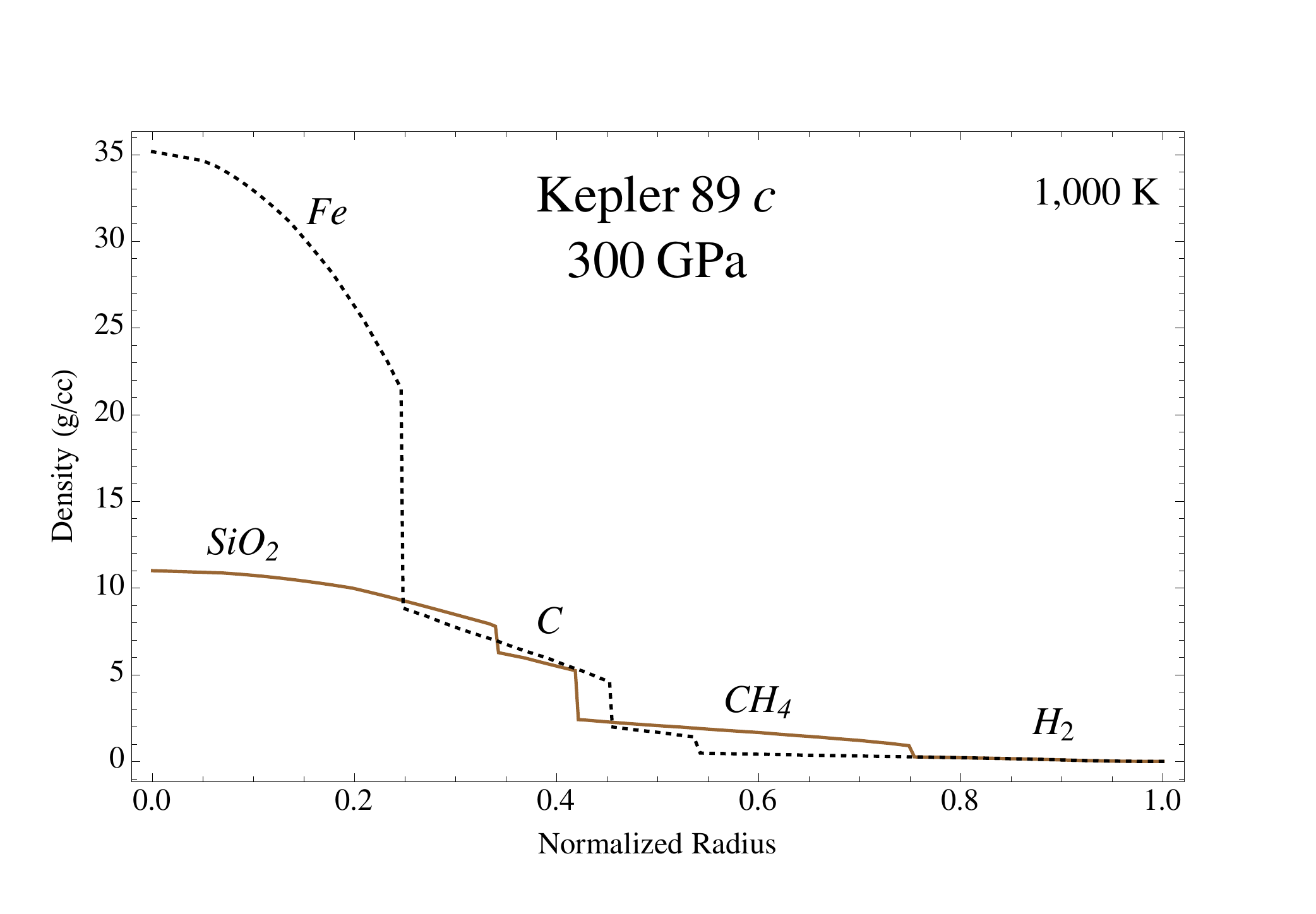}
    \label{fig:2}
  \end{subfigure}
  \caption{
  {\footnotesize {\bf Left:} Density vs.~normalized radius for  GJ3470 b with $P_{\rm diss}$=170 GPa (top), Kepler 89c with $P_{\rm diss}$=170 GPa (middle), and Kepler 89c with $P_{\rm diss}$=300 GPa (bottom).  The solid-black, dotted-gray, dashed-blue, and dashed-dotted-red curves correspond to temperatures of 50 K, 500 K, 1,000 K, and 1,500 K, respectively. 
{\bf Right}: Density vs.~normalized radius for GJ3470 b with $P_{\rm diss}$=170 GPa (top), Kepler 89c with $P_{\rm diss}$=170 GPa (middle) and Kepler 89c  with $P_{\rm diss}$=300 GPa (bottom) for  T = 1,000 K for two different core materials: SiO$_2$ (solid-brown) and Fe (dotted-black). As can be seen from the figure, an iron core leads to much higher internal densities. }
}
\end{figure}

\clearpage
\begin{table}[h!]
{\footnotesize
{\tiny
{\renewcommand{\arraystretch}{1.5}
\begin{tabular}{ | c c  c  c  c  c  c  c  c |}
\hline
\hline
{\bf Kepler 89c} & {\bf M=15.57$^{+15.57}_{-5.72}$M$_{\oplus}$} & {\bf R=4.33$\pm$0.09R$_{\oplus}$} & & & & & &\\
 \hline
 & M$_{\rm SiO_2}$/M$_p$ & R$_{\rm SiO_2}$/R$_p$ & M$_{\rm C}$/M$_p$ & R$_{\rm C}$/R$_p$  &  M$_{\rm CH_4}$/M$_p$ &  R$_{\rm CH_4}$/R$_p$ & M$_{\rm H_2}$/M$_p$ &  R$_{\rm H_2}$/R$_p$ \\
\hline
170 GPa & & & & & & & &\\
\hline
50 K & 0.001 & 0.1065 & 0.609 & 0.485 & 0.186 & 0.619 &0.203 &1.0\\
500 K & 0.21 & 0.291 & 0.458 & 0.484 & 0.179 & 0.617 & 0.153 & 1.0\\
1,000 K & 0.4 & 0.362 & 0.311 & 0.481 & 0.184 & 0.623 & 0.104 & 1.0\\
1,500 K &  0.51 & 0.395 & 0.230 & 0.4808 & 0.183 & 0.627 & 0.077 & 1.0\\
\hline
300 GPa & & & & & & & &\\
\hline
50 K & 2$\times 10^{-5}$ & 0.107 & 0.310 & 0.387 & 0.587 & 0.772 & 0.103 & 1.0\\
500 K & 0.12 & 0.251 & 0.264 & 0.403 & 0.528 & 0.753 & 0.088 & 1.0\\
1,000 K & 0.3 & 0.336 & 0.168 & 0.414 & 0.476 & 0.741 & 0.056 & 1.0\\
1,500 K & 0.39 & 0.3686 & 0.1098 & 0.4167 & 0.464 & 0.7434 & 0.0366  & 1.0\\
\hline
{\bf Kepler 20c} & {\bf M=15.7$\pm$3.3M$_{\oplus}$} & {\bf R=3.07$^{+0.2}_{-0.31}$R$_{\oplus}$} & & & & & &\\
 \hline
 & M$_{\rm SiO_2}$/M$_p$ & R$_{\rm SiO_2}$/R$_p$ & M$_{\rm C}$/M$_p$ & R$_{\rm C}$/R$_p$  &  M$_{\rm CH_4}$/M$_p$ &  R$_{\rm CH_4}$/R$_p$ & M$_{\rm H_2}$/M$_p$ &  R$_{\rm H_2}$/R$_p$ \\
 \hline
170 GPa & & & & & & & &\\
\hline
50 K & 0.74 & 0.647 & 0.055&  0.675 & 0.187 & 0.896 & 0.018 &1.0\\
500 K & 0.77 & 0.659 & 0.030 &  0.675 & 0.190 & 0.905 & 0.01 & 1.0\\
1,000 K & 0.79 & 0.668 & 0.011  & 0.673 & 0.196 & 0.921 & 0.003 & 1.0\\
1,500 K & 0.80 & 0.671 & 0.002 & 0.673 & 0.198 & 0.926 & 6$\times$10$^{-4}$ & 1.0\\
\hline
300 GPa & no solution & & & & & & &\\
\hline
300 GPa & no solution & & & & & & &\\
\hline
\hline
{\bf GJ 3470b} & {\bf  M=14.0$\pm$1.64M$_{\oplus}$} & {\bf  R=4.20$\pm$0.6R$_{\oplus}$} & & & & & &\\
\hline
 & M$_{\rm SiO_2}$/M$_p$ & R$_{\rm SiO_2}$/R$_p$ & M$_{\rm C}$/M$_p$ & R$_{\rm C}$/R$_p$  &  M$_{\rm CH_4}$/M$_p$ &  R$_{\rm CH_4}$/R$_p$ & M$_{\rm H_2}$/M$_p$ &  R$_{\rm H_2}$/R$_p$ \\
\hline
170 GPa & & & & & & & &\\
\hline
50 K & 0.01 & 0.132 & 0.564&  0.477 & 0.238 & 0.643 & 0.188 &1.0\\
500 K & 0.22 & 0.299 & 0.421  & 0.479 & 0.218 & 0.642 & 0.140 & 1.0\\
1,000 K & 0.42 & 0.373 & 0.267  & 0.475 & 0.224 & 0.646 & 0.089 & 1.0\\
1,500 K & 0.53 & 0.407 & 0.178 & 0.473 & 0.231 & 0.656 & 0.059 & 1.0\\
\hline
300 GPa & no solution & & & & & & &\\
\hline
\hline
\end{tabular} 
}
}
\caption{\label{exoplanets} 
{\footnotesize Solutions for a few detected exoplanets that are consistent with differentiated methane planets.  M$_{\rm SiO_2}$/M$_p$, M$_{\rm C}$/M$_p$, M$_{\rm CH_4}$/M$_p$,  M$_{\rm H_2}$/M$_p$ correspond to the mass fractions of SiO$_2$, C, CH$_4$ and H$_2$, respectively. R$_{\rm SiO_2}$/R$_p$, R$_{\rm C}$/R$_p$, R$_{\rm CH_4}$/R$_p$, and R$_{\rm H_2}$/R$_p$  correspond to the radii (normalized to the planet's radius) of the shells for the compositions of SiO$_2$, C, CH$_4$ and H$_2$, respectively. Given are the solutions for the calculated interior structure assuming dissociation pressures of 170 and 300 GPa, and four different internal temperatures: 50 K, 500 K, 1,000 K, and 1,500 K. The masses and radii of the planets are taken from the exoplanet database at http://exoplanets.org and are taken to be their 0$\sigma$ values. }
}
}
\end{table}

\begin{table}[h!]
{\footnotesize
\begin{center}
{\renewcommand{\arraystretch}{1.}
\begin{tabular}{ | c c  c  c  c |}
\hline
\hline
 \hline
{\bf Undifferentiated} & & & &\\
\hline
{\bf Kepler 11d}  &   {\bf M=6.1 $^{+3.1}_{-1.7}$ M$_{\oplus}$} & {\bf R=3.117 $^{+0.06}_{-0.07}$ R$_{\oplus}$}& &\\
\hline
 & M$_{\rm SiO_2}$/M$_p$ & R$_{\rm SiO_2}$/R$_p$ & M$_{\rm CH_4}$/M$_p$ &  R$_{\rm CH_4}$/R$_p$  \\
\hline
50 K  &  0.18 & 0.354  & 0.82 & 1.0 \\
500 K  &  0.22 & 0.376  & 0.78 & 1.0\\
1,000 K  & 0.28 &  0.403 & 0.72 & 1.0\\
1,500 K  & 0.33 &  0.423 & 0.67 & 1.0\\
\hline 
\hline
{\bf GJ 1214b}  & {\bf M=6.47 $\pm$ 1.0 M$_{\oplus}$} & {\bf  R=2.68 $\pm$ 0.13 R$_{\oplus}$}& &\\
\hline
 & M$_{\rm SiO_2}$/M$_p$ & R$_{\rm SiO_2}$/R$_p$ & M$_{\rm CH_4}$/M$_p$ &  R$_{\rm CH_4}$/R$_p$  \\
\hline
50 K  & 0.50  & 0.544  & 0.5 & 1.0 \\
500 K  & 0.53  & 0.554  & 0.47 & 1.0\\
1,000 K  & 0.58 &  0.570 & 0.42 & 1.0\\
1,500 K  & 0.62 &  0.583 & 0.38 & 1.0\\
\hline
\hline
{\bf HD 97658b}  & {\bf M=7.87 $\pm$ 0.73 M$_{\oplus}$} & {\bf R=2.34 $^{+0.18}_{-0.15}$ R$_{\oplus}$} & &\\
\hline
 & M$_{\rm SiO_2}$/M$_p$ & R$_{\rm SiO_2}$/R$_p$ & M$_{\rm CH_4}$/M$_p$ &  R$_{\rm CH_4}$/R$_p$  \\
\hline
50 K  & 0.88  & 0.811  & 0.12 & 1.0 \\
500 K  &  0.895 & 0.818  & 0.105  & 1.0\\
1,000 K  & 0.91 &  0.824 & 0.09 & 1.0\\
1,500 K  & 0.92 &  0.83 & 0.08 & 1.0\\
\hline
{\bf Kepler 68b}  & {\bf M=6.0 $\pm$ 1.71 M$_{\oplus}$} & {\bf R=2.308 $^{+0.06}_{-0.09}$ R$_{\oplus}$} & &\\
\hline
 & M$_{\rm SiO_2}$/M$_p$ & R$_{\rm SiO_2}$/R$_p$ & M$_{\rm CH_4}$/M$_p$ &  R$_{\rm CH_4}$/R$_p$  \\
\hline
50 K  &  0.85 & 0.782  & 0.15 & 1.0 \\
500 K  & 0.865  & 0.788  & 0.135 & 1.0\\
1,000 K  & 0.88 &  0.795 & 0.12 & 1.0\\
1,500 K  & 0.89 &  0.799 & 0.11 & 1.0\\
\hline
\hline
\end{tabular} 
}
\caption{\label{exoplanets}
{\footnotesize Solutions for exoplanets that potentially have methane-rich interiors but for which dissociation of methane does not occur.  Shown are the normalized masses and radii of the silicate (SiO$_2$) core and methane (CH$_4$) envelop for the four different internal temperatures used in this work. M$_{\rm SiO_2}$/M$_p$ and M$_{\rm CH_4}$/M$_p$ correspond to the mass fractions of SiO$_2$ and CH$_4$, respectively, while R$_{\rm SiO_2}$/R$_p$ andR$_{\rm CH_4}$/R$_p$ are the radii (normalized to the planet's radius) of the shells for the compositions of SiO$_2$, and CH$_4$, respectively. 
Also here, the masses and radii of the planets are taken from the exoplanet database at http://exoplanets.org and are taken to be their 0$\sigma$ values.}
}
\end{center} 
}
\end{table}

\end{document}